\documentclass[conference]{IEEEtran}
\IEEEoverridecommandlockouts
\usepackage{amsmath,amsfonts}

\usepackage{algorithm}
\usepackage{algorithmic}
\usepackage{multirow}
\usepackage{listings}%
\usepackage{tabularx}
\usepackage{textcomp}
\usepackage{stfloats}
\usepackage{url}
\usepackage{verbatim}
\usepackage{graphicx}
\usepackage{cite}
\usepackage{color}
\usepackage{amssymb}
\UseRawInputEncoding  
\usepackage{subcaption}
\usepackage{booktabs}
\usepackage{caption}
\usepackage{subcaption}

\begin{document}

\title{CGFormer: A Cross-Attention Based Grid-Free Transformer for Radio Map Estimation 
\thanks{This work was supported in part by the National Science Foundation grants \#2231209, \#2136202 and \#2128596.}
}
\author{\IEEEauthorblockN{Haihan Nan$^*$, \; Emmanuel Obeng Frimpong$^*$, \; Zhi Tian$^*$, \; Yue Wang$^\dagger$, \; Lingjia Liu$^\ddagger$,}
\IEEEauthorblockA{$^*$Department of Electrical and Computer Engineering, George Mason University, Fairfax, VA, USA\\$^\dagger$Department of Computer Science, Georgia State University, Atlanta, GA, USA\\$^\ddagger$ Department of Electrical and Computer Engineering, Virginia Tech, Blacksburg, VA, USA}}

\maketitle

\begin{abstract}
Radio map estimation (RME), which predicts wireless signal metrics at unmeasured locations from sparse measurements, has attracted growing attention as a key enabler of intelligent wireless networks. The majority of existing RME techniques employ grid-based strategies to process sparse measurements, where the pursuit of accuracy results in significant computational inefficiency and inflexibility for off-grid prediction. In contrast, grid-free approaches directly exploit coordinate features to capture location-specific spatial dependencies, enabling signal prediction at arbitrary locations without relying on predefined grids. However, current grid-free approaches demand substantial preprocessing overhead for constructing the spatial representation, leading to high complexity and constrained adaptability. To address these limitations, this paper proposes a novel cross-attention grid-free based transformer model for RME. We introduce a lightweight spatial embedding module that incorporates environmental knowledge into high-dimensional feature construction. A cross-attention transformer then models the spatial correlation between target and measurement points. The simulation results demonstrate that our proposed method reduces RMSE by up to 6\%, outperforming grid-based and grid-free baselines.

\end{abstract}


\section{Introduction}
Radio map estimation (RME) has emerged as a key enabler for intelligent and context-aware wireless systems by providing spatial representations of performance indicators across geographical regions. By predicting radio frequency (RF) metrics such as received signal strength (RSS), power spectral density (PSD), and channel gain~\cite{Bi2019_WC_survey, Zhang2022_ICC_PSD, Yang2025_Arxiv_PSD}, RME supports a wide range of applications including wireless localization, network planning, and resource allocation~\cite{Huang2019_IoT_localization, Romero2024_TWC_planning, Pesko2014_KSII_resourceallocation}. These applications highlight the importance of accurate and efficient RME for enhancing the adaptability and robustness of next-generation wireless networks.

Existing RME methods can be generally categorized into
model-based and learning-based paradigms. Model-based approaches, such as path-loss models~\cite{Kurt2017_APM_pathloss}, provide physically interpretable predictions and can achieve reasonable accuracy under fixed scenarios. However, their performance is often constrained by oversimplified propagation assumptions that fail to capture the variability of real-world environments. In addition, most model-based methods rely on comprehensive prior information, including transmitter power, node positions, and antenna orientations, which is rarely available in dynamic and heterogeneous wireless systems~\cite{Rautiainen2002_VTC_prior}. 

In contrast, learning-based methods can directly
learn propagation characteristics from measurements without
relying on explicit prior models~\cite{Feng2025_Elec_RMEsurvey}. Particularly, recent advances such as convolutional neural networks (CNNs) and vision transformers have shown remarkable performance in grid-based RME~\cite{Teganya2022_TWC_DCAE, Qiu2023_ICASSP, Bakirtzis2025_ICASSP}, where the region of interest is divided into uniform grids and radio metrics are predicted cell by cell. However, the grid-based formulation limits flexibility in representing irregular locations and requires processing the entire grid even when only sparse measurements or a small set of target points are involved, thereby incurring considerable computational cost. 

To solve this problem, grid-free methods have been proposed to overcome the discretization constraints of grid-based RME by directly modeling spatial representation at arbitrary location. Early grid-free estimators relied on classical interpolation techniques for RME, such as polynomial fitting or kernel-based regression~\cite{Bazerque2011_TSP_gridfree, Chen2025_SPL_gridfree}, which is inefficient in environments with sharp signal
transitions or irregular spatial variations. More recently, some deep learning-based frameworks have been introduced to enhance grid-free estimation. For instance, a transformer-based architecture named STORM~\cite{viet2024_arxiv_transformer} constructs translation- and
rotation-invariant spatial feature for each target point based on all measurements and utilize self-attention mechanism to capture spatial correlation among queries and measurements, demonstrating superior accuracy over traditional interpolators without relying on predefined
grids. However, such designs also suffer from notable drawbacks. First, constructing target-specific features requires translating all measurement coordinates with respect to each target location and performing value-dependent rotations, which leads to quadratic attention complexity in the number of measurement. Second, existing grid-free learning models typically focus on re-weighting measurement contributions but do not explicitly incorporate environmental semantics (e.g., building layouts or obstructions), reducing adaptability in heterogeneous propagation conditions. Third, self-attention mechanism is employed to represent the target point, which is redundant and lacks a clear query-conditioning structure, since the target embedding is entangled with all measurements in a symmetric manner. 

To address the aforementioned challenges, we propose a novel cross-attention grid-free based transformer (CGFormer) for RME. Specifically, we  first employs a spatial semantic embedding (SSE) module to encode raw coordinates through NeRF-style sinusoidal encoding scheme and integrate environmental priors, efficiently generating expressive target-centric spatial representations. Subsequently, a cross-attention based transformer model is adopted to capture the spatial correlation of the measurements on the targ5et points, where the target point acts as a query attending to the measurements (keys/values). In this way, our proposed framework not only reduces computational overhead but also provides a more interpretable and physically consistent representation. The main contributions of this paper are summarized as follows:

\begin{enumerate}
\item We propose CGFormer, a cross-attention based grid-free transformer framework tailored for radio map estimation. Unlike existing grid-based models that are tied to fixed resolutions or grid-free models that require excessive pre-processing operations, CGFormer directly models target points through flexible coordinate embeddings and cross-attention mechanism, thus enabling multi-resolution and adaptive deployment.
 
\item We design a spatial semantic embedding (SSE) module that encodes raw coordinates using a NeRF-style sinusoidal scheme and incorporates environmental priors, which efficiently captures fine-grained spatial semantics and provides expressive, target-centric representations for spatial correlation modeling.

\item The comprehensive experiments on Wireless Insite datasets
demonstrate that CGFormer consistently outperforms state-of-the-art grid-based and grid-free baselines in terms of estimation accuracy and generalization across different spatial resolutions.

\end{enumerate}

\section{Problem Formulation and Model Preliminaries}
This section first formulates the RME problem, followed by contrasting the grid-based and grid-free paradigms to highlight their differences and limitations. We further provide a brief introduction to the transformer architecture, which serves as the foundation for our proposed framework.

\subsection{Problem formulation}
We consider a spatial RME problem over a geographical region $\mathcal{X}$, where the objective is to estimate fine-grained spatial distributions of RF metrics from sparse sensor measurements. The RF metric at each location is inherently influenced by multiple factors such as the transmit power, the number of transmitters, temporal variations, and the operating frequency band. In this work, we restrict our study to a time-invariant single-band setting with two transmitters, while allowing their positions and powers to vary across scenarios. 

Specifically, the measurement set is defined as     $\mathcal{M}\triangleq\left\lbrace \left(\mathbf{p}^{\left(m\right)}_n, \tilde{\gamma}\left(\mathbf{p}^{\left(m\right)}_n\right)\right)\right\rbrace_{n=1}^{N}$, where $N$ denotes the number of measurements, $\mathbf{p}^{(m)}_n \in \mathcal{X}$ is the coordinate of the $n$-th measurement, and $\tilde{\gamma}\left(\mathbf{p}^{(m)}_n\right)$ is the corresponding observed metric value. The RME problem aims to estimate the RF metric $\hat{\gamma}\left(\mathbf{p}\right)$ at arbitrary location $\mathbf{p}\in\mathcal{X}$, given the measurement set $\mathcal{M}$ and optional environmental information $\mathcal{E}$ (e.g., building occupancy map $\mathbf{B}$ or sampling mask $\mathbf{S}$), which can be expressed as:
\begin{equation}
    \hat\gamma\left({\mathbf{p}}\right) = f\left(\mathbf{p},\mathcal{M}, \mathcal{E}\right),
\end{equation}
where $f(\cdot)$ denotes the estimator. If no side information is available, $f(\mathbf{p}, \mathcal{M}, \mathcal{E})$ naturally reduces to $f(\mathbf{p}, \mathcal{M})$.

\begin{figure}[t]
	\centering
	\subfloat[]{\includegraphics[width=1\linewidth]{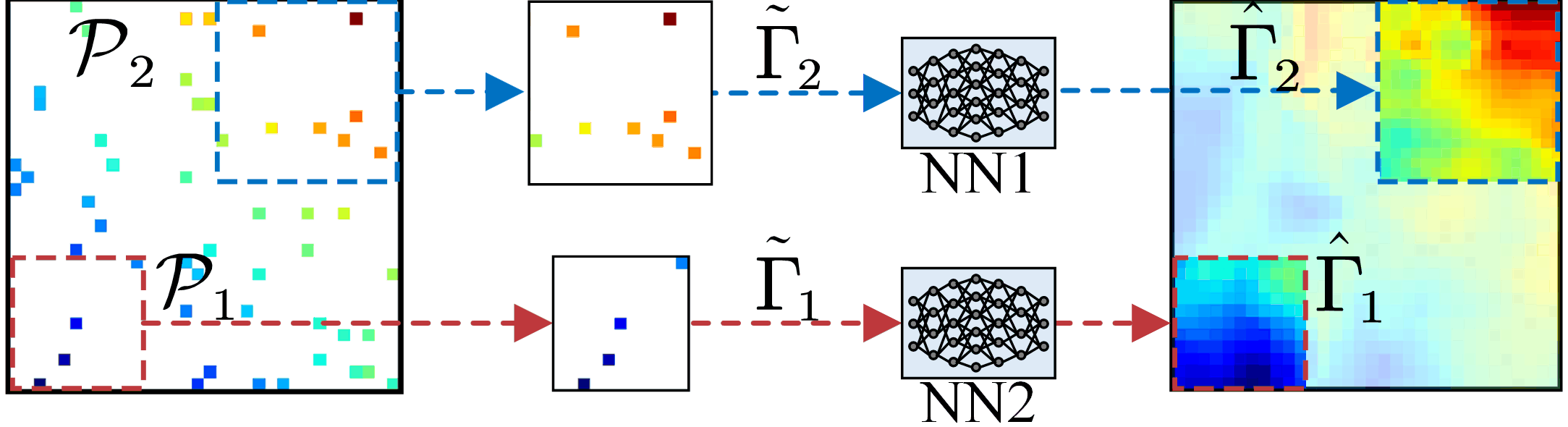}}\\
	\subfloat[]{\includegraphics[width=1\linewidth]{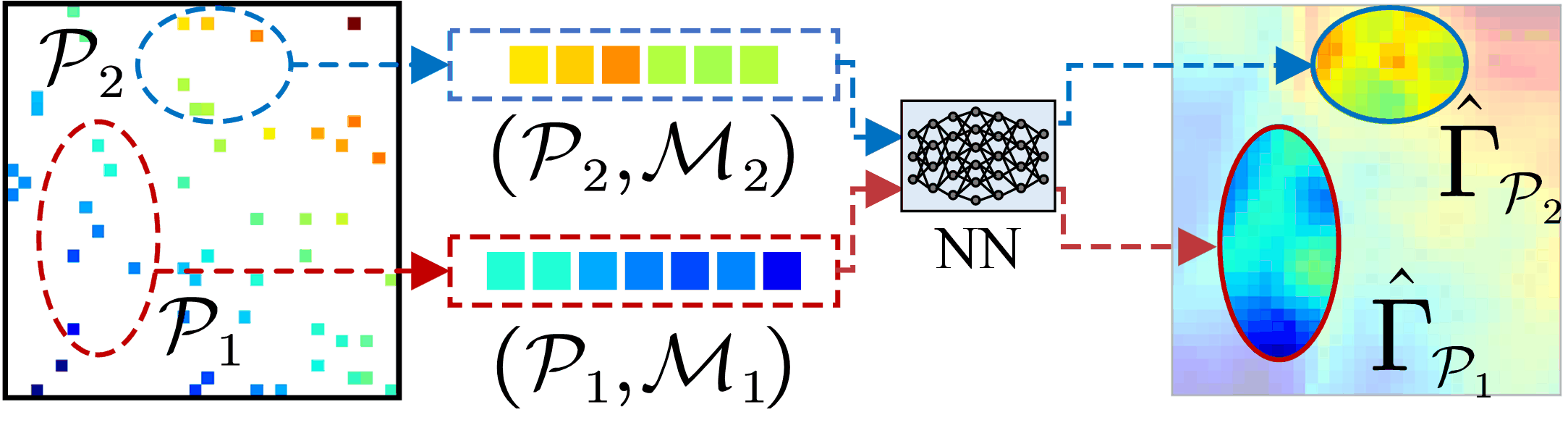}}
	\caption{Comparison of different RME paradigms. (a) grid-based RME, where the region is discretized into fixed grids, and (b) grid-free RME, where predictions are made at arbitrary spatial locations. External side information is not incorporated.}
	\label{fig1_rme}
\end{figure}

\subsection{Grid-based and Grid-free paradigms for RME}
As illustrated in Fig.~\ref{fig1_rme}, RME methods can be broadly categorized into grid-based and grid-free paradigms.  Grid-based schemes provide a simple formulation with fixed $N_y \times N_x$ complexity, but discretization restricts their flexibility in handling irregular locations and requires retraining for different resolutions. In contrast, grid-free RME allows inference at arbitrary coordinates without predefined grids, avoiding discretization errors and better preserving fine-grained spatial variations. It further supports multi-resolution and heterogeneous tasks, as the query set $\mathcal{P}$ can be adaptively specified without model retraining.

In the grid-based paradigm, the region $\mathcal{X}$ is discretized into an $N_y \times N_x$ rectangular grid with intervals $\Delta_{y}$ and $\Delta_{x}$. The grid cell $\left(i, j\right)$ is located at $\boldsymbol{\xi}_{i,j} \triangleq [j\Delta{x}, i\Delta_{y}]^{\mathrm{T}}$, where $i \in \{1,...,N_y\}, j \in \{1,...,N_x\}$. Each cell aggregates and averages all measurements within its coverage to obtain the sampled value, which is defined as:
\begin{equation}
\tilde{\gamma}\left(\boldsymbol{\xi}_{i,j}\right) \!=\! \frac{1}{\left|N_{i,j}\right|}\!\sum\nolimits_{n=1}^N{\tilde{\gamma}\left( \mathbf{p}_{n}^{\left( m \right)} \right) \mathbb{I}\left( \mathbf{p}_{n}^{\left( m \right)}\!\in\! \mathcal{N}\left(\boldsymbol{\xi}_{i,j}\right) \right)},
\end{equation}
where $\mathcal{N}(\boldsymbol{\xi}_{i,j}) 
\!\!=\!\! \left\{ \mathbf{p} \;\middle|\; 
\|\mathbf{p}-\boldsymbol{\xi}_{i,j}\| \!\leq \!\|\mathbf{p}-\boldsymbol{\xi}_{k,l}\|, \;\forall (k,l) \!\neq\! (i,j) \right\}$ is the set of measurement coordinates mapped to grid cell $\left(i,j\right)$ and $N_{i,j}$ is the number of measurements at corresponding cell, and $\tilde{\gamma}(\boldsymbol{\xi}_{i,j})$ denote the measured value at $\boldsymbol{\xi}_{i,j}$, which is set to 0 if the cell has no measurements.

By aggregating measurements across all grid cells, the grid-based representation is given by $\boldsymbol{\tilde\Gamma}=\left[\tilde\gamma\left(\boldsymbol{\xi}_{i,j}\right)\right]_{i=1,j=1}^{N_y,N_x}\in \mathbb{R}^{N_y\times N_x}$. The ground-truth RF distribution is similarly expressed as $\boldsymbol{\Gamma}^{\ast}=\left[\gamma^{\ast}\left(\boldsymbol{\xi}_{i,j}\right)\right]_{i=1,j=1}^{N_y,N_x}\in \mathbb{R}^{N_y\times N_x}$, where $\gamma^{\ast}\left(\boldsymbol{\xi}_{i,j}\right)$ represents the true value at $\boldsymbol{\xi}_{i,j}$. Side information, such as building maps or sampling masks, can also be encoded on the same grid, denoted by $\mathbf{B}\in \mathbb{R}^{N_y\times N_x}$ and $\mathbf{S}\in \mathbb{R}^{N_y\times N_x}$. Given $\boldsymbol{\tilde\Gamma}$ and optional side information $\mathcal{E}\subseteq\{\mathbf{B},\mathbf{S}\}$, the grid-based RME task is to recover RF values over all grids in $\mathcal{X}$:
\begin{equation}
\hat{\boldsymbol{\Gamma}}=f\left(\boldsymbol{\tilde\Gamma}, \mathcal{E}\right),
\end{equation}
where $\hat{\boldsymbol{\Gamma}}$ is the predicted RF map and $\mathcal{E}$ denotes auxiliary information (e.g., $\mathbf{B}$ and $\mathbf{S}$) represented either as separate matrices or a unified composite form~\cite{Teganya2022_TWC_DCAE}.

In contrast to grid-based RME, grid-free RME directly infers metric values at arbitrary locations without discretizing $\mathcal{X}$. For a query set $\mathcal{P}=\{\mathbf{p}_1,\dots,\mathbf{p}_Q\}$, the predictions are expressed as $\hat{\boldsymbol{\Gamma}}^{\mathcal{P}}=\big[\hat{\gamma}(\mathbf{p}_1), \hat{\gamma}(\mathbf{p}_2), \dots, \hat{\gamma}(\mathbf{p}_Q)\big]^{\mathrm{T}}
= f\left(\mathcal{P}, \mathcal{M}, \mathcal{E}\right)$, where ${\boldsymbol{\hat\Gamma}}^{\mathcal{P}}$ denotes the predicted values at $\mathcal{P}$.

\subsection{Encoder-Style Transformer Blocks} 
Given the limitations of grid-based CNNs and the flexibility of grid-free formulations, we adopt a Transformer-inspired design to model global spatial dependencies in RME. Rather than employing the full encoder-decoder Transformer, our model comprises a stack of $N_l$ encoder-style attention blocks, each including a multi-head self-attention (MHA) layer and a position-wise feed-forward network, with residual connections and pre-layer normalization to ensure stable training.

Given query $\mathbf{Q}\in\mathbb{R}^{N_Q\times d_Q}$, key $\mathbf{K}\in\mathbb{R}^{N_K\times d_K}$, and value $\mathbf{V}\in\mathbb{R}^{N_V\times d_V}$ sequences, the MHA operator is defined as: \begin{subequations}
    \begin{align}
        \mathrm{MHA}\left(\mathbf{Q}, \mathbf{K}, \mathbf{V}\right) = \left[\mathrm{head}_1,...,\mathrm{head}_h\right]\mathbf{W}^O,\\
        \mathrm{head}_r=\textrm{softmax}\left(\frac{\left(\mathbf{Q}\mathbf{W}^Q_r\right)\left(\mathbf{K}\mathbf{W}^K_r\right)^T}{\sqrt{d_h}}\right)\mathbf{V}\mathbf{W}^V_r,
    \end{align}
\end{subequations}
where $h$ is the number of heads, $d_h=d/h$ is the dimensionality of each head, and $\mathbf{W}^Q_r,\mathbf{W}^K_r,\mathbf{W}^V_r,\mathbf{W}^O$ are projection matrices of query, key, value, and output, respectively. Each block then applies the MHA output through a feed-forward module $f_\text{MLP}(Z)=\mathbf{W}_2\phi(\mathbf{W}_1Z)$, denoted as:
\begin{subequations}
     \begin{align}
        \mathbf{Q}^{\prime} = \mathbf{Q} + \mathrm{MHA}\left( l_1\left( \mathbf{Q} \right) , \mathbf{K}, \mathbf{V} \right) , \\
        T\left( \mathbf{Q}, \mathbf{K}, \mathbf{V} \right) = \mathbf{Q}^{\prime}+f_\textrm{MLP}\left( l_2\left( \mathbf{Q}^{\prime} \right) \right),
    \end{align}
    \label{eq_6b}
\end{subequations}
\leavevmode\hspace*{-\parindent} with independent LayerNorms $l_1$,$l_2$. Stacking such blocks allows transformers to capture long-range dependencies and refine contextualized representations.

\section{Proposed method}
In this section, we present the proposed CGFormer framework for grid-free radio map estimation. We first provide an overview of the overall framework, followed by detailed descriptions of its key components, including the spatial semantic embedding (SSE) module and the cross-attention transformer prediction module.


\begin{figure}
    \centering
    \includegraphics[width=0.9\linewidth]{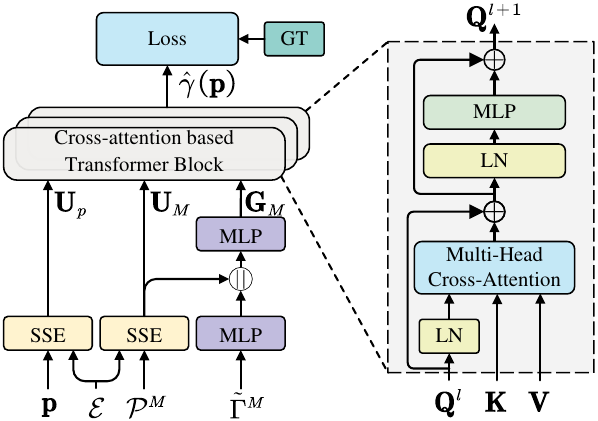}
    \caption{Overall architecture of the proposed CGFormer framework.}
    \label{fig:CGFormer}
\end{figure}

\subsection{Overall framework}
As illustrated in Fig.~\ref{fig:CGFormer}, the proposed CGFormer framework estimates RF signal strengths at arbitrary target locations from sparse measurements and environmental priors. The input includes target coordinates $\mathbf{p}$, environmental variables $\mathcal{E}$, measurement coordinates $\mathcal{P}^M$, and corresponding measurement values $\mathbf{\tilde{\Gamma}}^M$. All coordinates, together with $\mathcal{E}$, are first encoded by the SSE module, which produces spatial features for target and measurement points. The resulting target features $\mathbf{U}_p$ serve as queries, while the measurement features $\mathbf{U}_M$ act as keys in the cross-attention enhanced prediction block. Meanwhile, measurement values $\mathbf{\tilde{\Gamma}}^M$ are transformed through a two-stage MLP and concatenated with $\mathbf{U}_M$ to obtain the value features $\mathbf{G}_M$, which can be expressed as:
\begin{equation}
    \mathbf{G}_M=\mathrm{MLP}_{\theta_2}\left(\left[\mathrm{MLP}_{\theta_1}\!\!\left(\mathbf{\tilde\Gamma}^M\right)\Vert \mathbf{U}_M\right]\right),
\end{equation}
where $\mathrm{MLP}_{\theta_1}$ and $\mathrm{MLP}_{\theta_2}$ are two-layer MLP with 64 and 32 hidden units. This design allows the measurement values to be effectively fused with their positional information, providing a compact and expressive representation.

Finally, multiple cross-attention based transformer blocks iteratively update the queries by attending to measurement features, yielding predictions $\hat\gamma\left(\mathbf{p}\right)$. Model training is supervised by minimizing the RMSE loss between predicted and ground-truth signal values.

\subsection{Spatial semantic embedding module}
\begin{figure}
    \centering
    \includegraphics[width=1\linewidth]{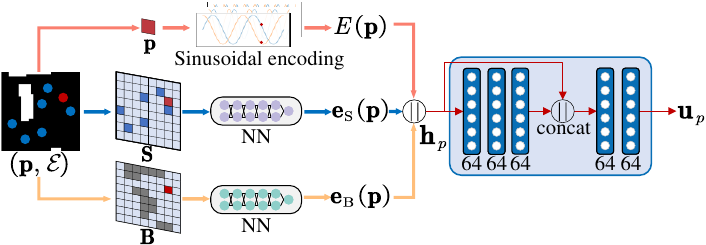}
    \caption{Architecture of the SSE module, which encodes spatial coordinates $\mathbf{p}$ and environmental priors $\mathcal{E}$ into NeRF-style sinusoid coordinate embeddings $E\left(\mathbf{p}\right)$ and global semantic embeddings $\mathbf{e}_\mathrm{B}\left(\mathbf{p}\right)$ and $\mathbf{e}_\mathrm{S}\left(\mathbf{p}\right)$ for buildings and measurements, respectively.}
    \label{fig:sse}
\end{figure}

The SSE module enhances spatial representations by encoding raw coordinates with a NeRF-style sinusoidal mapping~\cite{Mildenhall2021_ACMC_Nerf, Zhao2023_Nerf2}, which captures high-frequency variations of wireless propagation beyond the information contained in raw 2D positions. As shown in Fig.~\ref{fig:sse}, each point $\mathbf{p}\in\mathcal{P}$ or $\mathcal{P}^M$ is transformed through sinusoidal encoding $E(\mathbf{p})$, yielding richer embeddings that facilitate modeling multipath effects and spatial irregularities. The encoding function is defined as:

\begin{equation}
E(x) = \big[x,\; \{ \sin(2^k \pi x), \cos(2^k \pi x) \}_{k=0}^{L-1} \big],
\end{equation}
\noindent where the function $E\left(\cdot\right)$ is separately applied to each of the two coordinate values in $\mathbf{p}$. 

In addition, environmental context is incorporated through semantic priors such as building maps $\mathbf{B}$ and measurement masks $\mathbf{S}$. Both priors are represented as 2D grids of size $N_y\times N_x$, aligned with the underlying spatial layout. To transform these raw binary maps into expressive features, we employ a lightweight three-layer convolutional encoder. Formally, for each prior $\mathbf{X}\in\{\mathbf{B},\mathbf{S}\}$, the encoder is defined as:
\begin{equation}
\begin{aligned}
\mathbf{E}_x&=f_{\mathrm{CNN}}\left(\mathbf{X}\right)\\
&=\mathrm{Conv}_{3}\left(\mathrm{ReLU}\left(\mathrm{Conv_{3}\left(\mathrm{ReLU}\left(\mathrm{Conv_{5}}\left(\mathbf{X}\right)\right)\right)}\right)\right),
\end{aligned}
\end{equation}
where $\mathrm{Conv}_k$ denotes a 2D convolution with kernel size $k$, stride 1, and zero-padding to preserve the spatial resolution, and ReLU is used as the activation function. This design is computationally lightweight yet effective in extracting structural information such as building boundaries and measurement accessibility patterns. For each coordinate $\mathbf{p}$, we retrieve its nearest-grid semantic vectors $\mathbf{e}_\mathrm{B}\left(\mathbf{p}\right)$ and $\mathbf{e}_\mathrm{S}\left(\mathbf{p}\right)$ from the corresponding feature maps. The semantic features are concatenated with sinusoid-encoded coordinate embedding $E\left(\mathbf{p}\right)$ to form the target-centric embedding:
\begin{equation}
    \mathbf{h}_p=\left[ \left. E\left( \mathbf{p} \right) \right\| \left. \mathbf{e}_{\mathrm{B}}\left( \mathbf{p} \right) \right\| \mathbf{e}_{\mathrm{S}}\left( \mathbf{p} \right) \right]\in \mathbb{R}^{2\times \left(2L+1\right)+d_b+d_s},
\end{equation}
where $d_b$ and $d_s$ denote the dimensions of $\mathbf{e}_\mathrm{B}\left(\mathbf{p}\right)$ and $\mathbf{e}_\mathrm{S}\left(\mathbf{p}\right)$, respectively.

To further refine the latent representation $\mathbf{h}_p$, we employ a residual MLP $\mathrm{MLP}_{\theta}\left(\mathbf{h}_p\right)$, which consists of 5 fully connected layers with ReLU activations and a skip connection at the third layer. The residual design facilitates stable training and effective feature transformation while preserving the original information in $\mathbf{h}_p$. The output of $\mathrm{MLP}_{\theta}\left(\cdot\right)$
is a high-dimensional feature vector $\mathbf{u}_p$ that serves as the refined representation for subsequent prediction.

\subsection{Cross-attention enhanced prediction module}
Cross-attention enhanced prediction module aims
to capture spatial dependencies between queries and available measurement. Given the spatial features, we construct the transformer inputs as queries $\mathbf{Q}^{\left(0\right)}=\mathbf{U}_\mathcal{P}=\left[\mathbf{u}_1,\mathbf{u}_2,...,\mathbf{u}_Q\right]$, keys $\mathbf{K}=\mathbf{U}_M=\left[\mathbf{u}^{\left(m\right)}_1,\mathbf{u}^{\left(m\right)}_2,...,\mathbf{u}^{\left(m\right)}_N\right]$, and values $\mathbf{V}=\mathbf{G}_M$. Within each transformer block, multi-head cross-attention updates the query features by attending to the measurement features, followed by normalization and feed-forward layers. At each layer $l$, the query representations are updated through the transformer operator with $N_h$ heads:
\begin{equation}
\mathbf{Q}^{\left(l+1\right)}=T\left(\mathbf{Q}^{\left(l\right)}, \mathbf{K}, \mathbf{V}\right).
\end{equation}
After $L_t$ layers, the final queries $\mathbf{Q}^{\left(L_t\right)}$ are projected through a linear mapping $\ell(\cdot)$ to yield the estimated RF metric values, i.e., $\hat{\Gamma}_\mathcal{P}=\ell\left(\mathbf{Q}^{\left(L_t\right)}\right)\in \mathbb{R}^{Q}$. This iterative update mechanism enables the queries to adaptively incorporate contextual information from the measurement features, thereby enhancing the prediction accuracy. 

For model training, we adopt the mean squared error (MSE) as the objective function, which directly measures the discrepancy between predicted and ground-truth RF values:
\begin{equation}
\mathcal{L}_\mathrm{MSE}=\frac{1}{Q}\|\hat{\boldsymbol\Gamma}_\mathcal{P}-{\boldsymbol\Gamma}^{\ast}_\mathcal{P}\|_2^2
\end{equation}
where $\hat{\boldsymbol\Gamma}_\mathcal{P}$ and ${\boldsymbol\Gamma}^{\ast}_\mathcal{P}$ denote the predicted and ground-truth signal values at the Q query points, and $\|\cdot\|$ denotes the Euclidean norm.

\section{Experimental Results}
This section presents the experimental evaluation of the proposed CGFormer framework. The datasets and setup are first introduced, followed by performance comparisons with state-of-the-art baselines. We then present ablation studies to analyze the impact of key modules and demonstrate the framework’s adaptability across different spatial resolutions.

\subsection{Experimental setting}
\subsubsection{Dataset}
We evaluate CGFormer on the quasi-realistic Wireless InSite dataset. The raw datasets are generated using the Remcom Wireless InSite ray-tracing simulator in an urban canyon scenario, which captures multipath propagation and obstacle interactions. A total of 42 raw power maps are obtained, each corresponding to a single transmitter deployment. The original maps with size $656.1~\mathrm{m}\times 661.5~\mathrm{m}$ are uniformly discretized into $244 \times 246$ grids (i.e., spatial interval $\Delta=3.25~\mathrm{m}$) with two key attributes, including 2D coordinates and received power. The training maps are built by randomly aggregating two maps from the index set $\mathcal{I}_{\mathrm{train}}=\left\{1,2,...,33\right\}$. For each aggregated map, a random $16 \times 16$ sub-region is extracted, and the sampling factor is uniformly drawn from $\left[0.04,0.8\right]$. To simulate the practical constraint that only partial measurements are available, we adopt the strategy proposed in prior work \cite{Teganya2022_TWC_DCAE}. Specifically, each $16\times 16$ sub-region is further augmented into 10 sub-samples by randomly splitting the measurements into two subsets: 50\% are retained as observed inputs and the remaining 50\% are used as prediction targets. Based on the above construction, we generate $1.25\times10^6$  samples, which are further divided into training and validation sets with a 0.9 and 0.1 split. Similarly, the independent test set is derived from the raw maps indexed by $\mathcal{I}_{\mathrm{test}}=\left\{34,...,42\right\}$, from which 1000 aggregated maps are generated with two transmitters, ensuring no overlap with the training data.

\subsubsection{Baselines} 
We compare CGFormer with a diverse set of baselines, including KNN \cite{Larose2014_KNN}, Kriging~\cite{Cressie1988_Kriging}, Gaussian Process Regression (GPR)~\cite{Rasmussen2004_GPR}, and deep learning methods. For the latter, we consider both grid-based methods such as the Deep Completion Autoencoder (DCAE)~\cite{Teganya2022_TWC_DCAE} and grid-free methods such as STORM~\cite{viet2024_arxiv_transformer}, a transformer-based model with translation- and rotation-invariant spatial features.

\subsubsection{Evaluation metrics}
The prediction performance is evaluated using a widely adopted error metric: the root mean square error (RMSE), which is defined as:
\begin{equation}
    \mathrm{RMSE} = \sqrt{ \frac{1}{N_t} \sum_{i=1}^{N_t} \left( \hat{\gamma}_i - \gamma_i \right)^2 },
\end{equation}
where $\hat{\gamma}_i$ and ${\gamma}_i$ denote the predicted and ground-truth signal strength at the $i$-th location, respectively. $N_t$ represents the total number of available test points sampled from non-building regions of the radio map.

\begin{figure}
    \centering
    \includegraphics[width=1\linewidth]{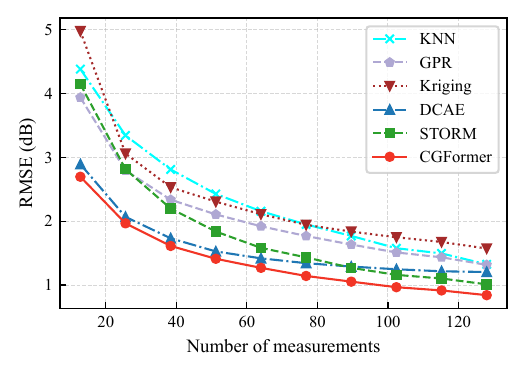}
    \caption{RMSE comparison of different methods under varying numbers of measurements on the Wireless InSite dataset.}
    \label{fig:rmse}
    \vspace{-0.1in}
\end{figure}

\subsubsection{Implementation details} Raw coordinates are mapped into a high-dimensional space with $L=16$, $d_b=16$, $d_s=16$ in the SSE modules. Meanwhile, the residual MLP is implemented with 64 hidden units, producing feature embeddings of 32 dimensions. The cross-attention enhanced module adopts a lightweight transformer backbone with $L_t=2$ stacked block, each using a hidden size of 64 for all vectors and $N_h=4$ attention heads. we employ Adam optimizer with an initial learning rate of $5\times10^{-4}$ and a batch size of 64. Models are trained for 200 epochs on an NVIDIA 3090 GPU with early stopping based on validation loss.

\subsection{Experimental results}
\subsubsection{Baseline comparison}
To evaluate prediction accuracy, we compare the RMSE of different baselines on the Wireless InSite datasets under different sampling factors. The test sampling factor is varied from 0.05 to 0.5 in increments of 0.05, simulating scenarios with varying levels of measurement availability.

As shown in Fig~\ref{fig:rmse}, deep learning-based methods consistently outperform classical interpolation and kernel-based baselines across most sampling factors, highlighting their advantages in capturing complex propagation patterns. Meanwhile, Both CGFormer and STORM leverage transformer architecture to model spatial correlation, which demonstrate notable performance gains at higher sampling factors, whereas the grid-based DCAE shows slower improvement. Overall, CGFormer achieves the best performance across all sampling factors, reducing RMSE by an average of 12.9\% compared with the second-best DCAE baseline.

\begin{figure}
    \centering
    \includegraphics[width=1\linewidth]{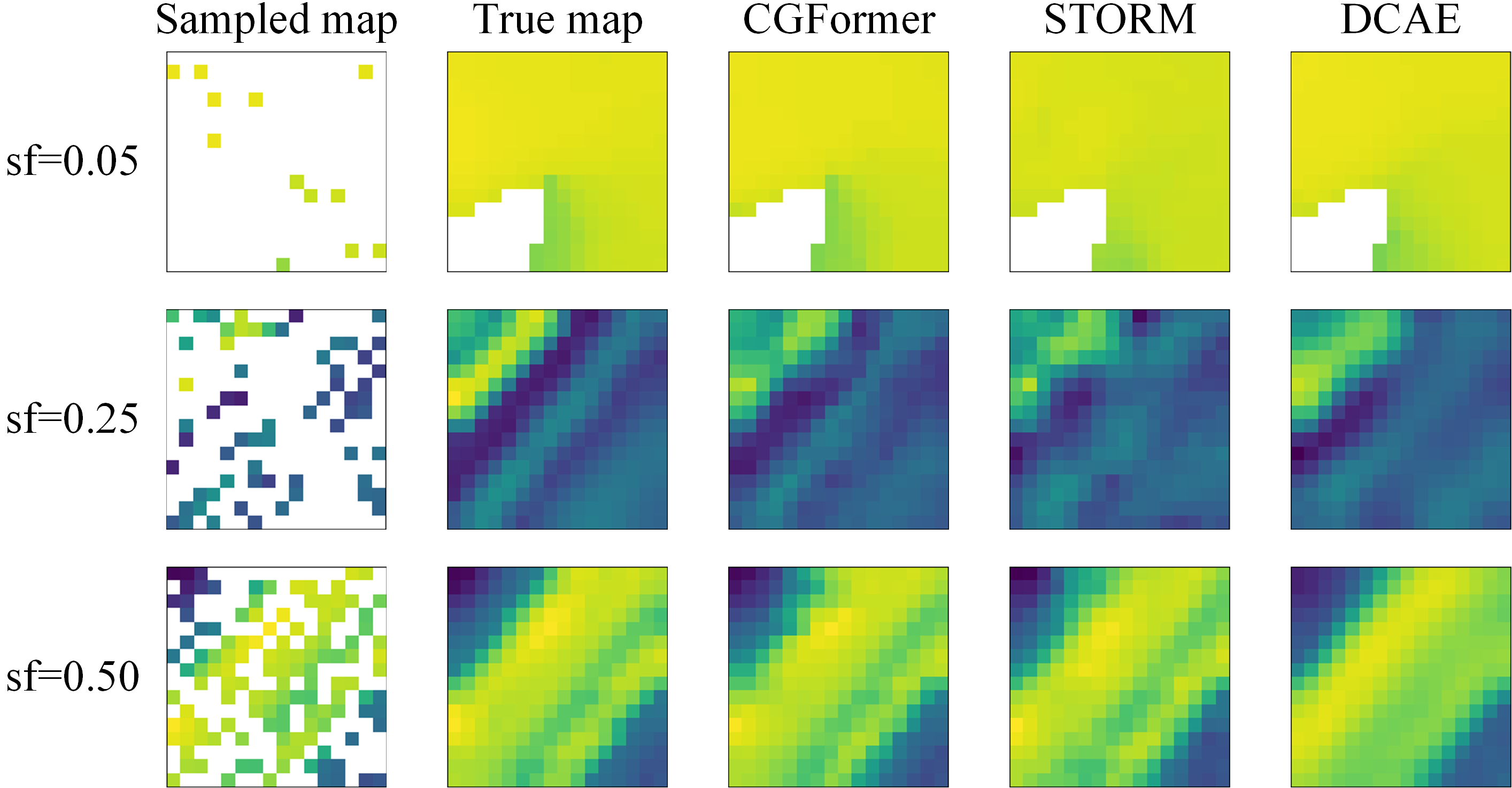}
    \caption{Visualization of sampled, ground-truth, and estimated maps produced by CGFormer, STORM, and DCAE under sampling factors of 0.05, 0.25, and 0.5. White regions correspond to building areas.}
    \label{fig:visual}
\end{figure}

To more intuitively illustrate the superiority of CGFormer for RME, we randomly visualize the true map and reconstructed maps of different methods under three different sampling factors. As illustrated in~\ref{fig:visual}, CGFormer can reconstruct radio maps more accurately than STORM and DCAE across all sampling factors, effectively capturing both high and low signal strength regions. While STORM improves under higher sampling rates, its performance degrades significantly with sparse measurements. In contrast, DCAE produces overly smoothed estimates and struggles to model sharp signal boundaries.

\subsubsection{Ablation study}
To evaluate the contribution of each component in the SSE module, we conduct ablation studies by selectively removing positional encoding, building map encoding, and measurement mask encoding. Specifically, we compare the following variants: CGFormer without sinusoidal positional encoding (CGFormer w/o PosEnc), CGFormer without building map encoding (CGFormer w/o $\mathbf{B}$), and CGFormer without measurement mask encoding (CGFormer w/o $\mathbf{S}$). 

As shown in Table~\ref{tab:ablation}, the performance will degrade when removing any component in the SSE module, confirming their effectiveness in enhancing spatial representation. Among them, positional encoding has the most significant impact on RMSE, indicating its critical role in capturing high-frequency variations in wireless propagation. The inclusion of building maps and measurement masks further improves accuracy by integrating semantic priors, but their contributions are relatively smaller. Overall, the full CGFormer model, which combines all three components, achieves the lowest RMSE and delivers the best performance.

\subsubsection{Adaptivity to off-grid query}
\begin{figure}
    \centering
    \includegraphics[width=0.7\linewidth]{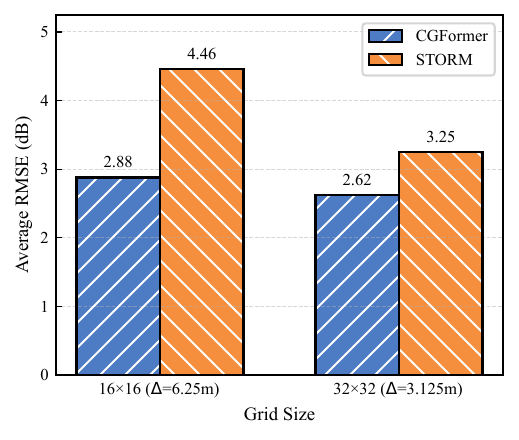}
    \caption{Average RMSE of CGFormer and STORM under different grid sizes, corresponding to spatial intervals $\Delta$ of 6.25 m and 3.125 m, respectively.}
    \label{fig:off-grid}
\end{figure}

Practical deployments often involve heterogeneous spatial resolutions and irregular query locations. To assess adaptivity, we evaluate CGFormer in a representative off-grid prediction scenarios. The model is trained on $16\times16$ maps ($\Delta=6.5 \mathrm{m}$) and directly tested on finer $32\times32$ maps ($\Delta=3.25 \mathrm{m}$). The finer grid introduces many off-grid query points unseen during training. As shown in Fig.~\ref{fig:off-grid}, CGFormer achieves significantly lower RMSE than STORM under both matched and unseen resolutions, demonstrating strong generalization beyond the training resolution. Notably, despite the larger fraction of off-grid queries, CGFormer benefits from denser measurements and achieves even higher accuracy at finer resolution. Overall, CGFormer adapts effectively to both varying spatial resolutions and unseen query points without retraining, underscoring its practicality for dynamic deployment conditions.

\begin{table}[t]
\centering
\caption{Ablation study on the SSE module. The comparison includes removing NeRF-style sinusoid positional encoding (PosEnc), building map ($\mathbf{B}$), and measurement mask ($\mathbf{S}$). The full CGFormer model integrates all three components.}
\label{tab:ablation}
\begin{tabular}{l c}
\toprule
\textbf{Variant} & \textbf{Average RMSE (dB)} \\
\midrule
\textbf{Full CGFormer} & 1.387 \\
w/o PosEnc & 1.655 \\
w/o $\mathbf{B}$ & 1.423 \\
w/o $\mathbf{S}$ & 1.416 \\
\bottomrule
\end{tabular}
\end{table}


\vspace{-0.01in}
\section{Conclusions}
\vspace{-0.02in}
This paper introduced CGFormer, a cross-attention based grid-free framework for radio map estimation. CGFormer integrates spatial semantic embeddings and an cross-attention enhanced prediction module to efficiently capture complex spatial correlations. Experimental results on quasi-realistic datasets show the effectiveness of CGFormer in delivering accurate and efficient radio map estimation compared with grid-based and grid-free baselines. Its capability to handle varying resolutions and off-grid predictions enables practical applications in the next-generation networks.

\vspace{-0.02in}
\bibliographystyle{IEEEtran}
\bibliography{bib}

\end{document}